\documentclass[12pt]{iopart}

\usepackage{cite}

\usepackage{graphicx}

\usepackage{dcolumn}

\begin{document}

\title[Conditionally solvable model]{Comment on:``Harmonic oscillator in an environment with a pointlike
defect''. Phys. Scr. \textbf{94} ( 2019) 125301 }
\author{Francisco M. Fern\'{a}ndez}

\address{INIFTA, Divisi\'on Qu\'imica Te\'orica,
Blvd. 113 S/N,  Sucursal 4, Casilla de Correo 16, 1900 La Plata,
Argentina}\ead{fernande@quimica.unlp.edu.ar}

\maketitle

\begin{abstract}
We analyze recent results for a harmonic oscillator in an environment with a
pointlike defect. We show that the allowed oscillator frequencies predicted
by the authors stem from a misinterpretation of the exact solutions of a
conditionally solvable eigenvalue equation. Also the exact eigenvalues
derived by those authors are meaningless because they belong to different
quantum-mechanical models.
\end{abstract}

\pacs{03.65.Ge}

In a recent paper Vit\'{o}ria and Belich\cite{VB19} investigated the
topology effects of the medium on a harmonic oscillator and the effects of a
Coulomb and linear central potentials on the harmonic oscillator in an
environment with a pointlike defect. The Schr\"{o}dinger equation is
separable in spherical coordinates and they solved the eigenvalue equation
for the radial part by means of the Frobenius method. Since the coefficients
of the expansion satisfy a three-term recurrence relation they could obtain
exact polynomial solutions and eigenvalues by truncation of the series. They
concluded that the angular frequency of the harmonic oscillator has
restricted values determined by the quantum numbers of the system. The
purpose of this comment is to analyze the effect of the truncation approach
on the results obtained by Vit\'{o}ria and Belich and on the physical
conclusions drawn from them.

The eigenvalue equation for the third model discussed by the authors (given
by the potential, $V(r)=k/r+m\omega ^{2}r^{2}/2$) is
\begin{eqnarray}
&&u^{\prime \prime }(s)+\frac{1}{s}u^{\prime }(s)-\frac{\iota ^{2}}{s^{2}}%
u(s)-\frac{\gamma }{s}u(s)-s^{2}u(s)+\delta u(s)=0,  \nonumber \\
&&\gamma =\frac{2k}{\left( \alpha \hbar \right) ^{3/2}}\sqrt{\frac{m}{\omega
}},\;\delta =\frac{2\mathcal{E}}{\alpha \hbar \omega },\;\iota ^{2}=\frac{%
4l(l+1)+\alpha ^{2}}{\alpha ^{2}},  \label{eq:eig_eq_VB_CH}
\end{eqnarray}
where $l=0,1,\ldots $ is the angular momentum quantum number, $m$ the mass
of the particle, $\alpha $ the parameter associated to the pointlike global
monopole and $\mathcal{E}$ the energy. By means of the truncation method
that we discuss below the authors obtained an expression for the energy $%
\mathcal{E}_{l,\bar{n}}=\alpha \hbar \omega _{l,\bar{n}}\left( 1+\bar{n}%
+|\iota |\right) $, where $\bar{n}=1,2,\ldots $ denotes the \textit{radial
modes} and $\omega _{l,\bar{n}}$ the \textit{allowed} angular frequency
that, according to the authors, depends on the quantum numbers. By
straightforward inspection one immediately suspects that something is amiss
here because the eigenvalue equation (\ref{eq:eig_eq_VB_CH}) exhibits bound
states for all $-\infty <\gamma <\infty $; therefore, there is no room for
such discrete values of $\omega $. One expects, of course, allowed values of
$\delta $ (or $\mathcal{E}$) that is the eigenvalue in this equation.

The fourth example comes from the potential $V(r)=\eta r+m\omega ^{2}r^{2}/2$
and the eigenvalue equation for the radial part is
\begin{eqnarray}
&&u^{\prime \prime }(s)+\frac{1}{s}u^{\prime }(s)-\frac{\iota ^{2}}{s^{2}}%
u(s)-\theta su(s)-s^{2}u(s)+\delta u(s)=0,  \nonumber \\
&&\theta =\frac{2\eta }{\sqrt{\alpha \hbar m\omega ^{3}}}.
\label{eq:eig_eq_VB_LH}
\end{eqnarray}
The authors obtained the exact energies $\mathcal{E}_{l,\bar{n}}=\alpha
\hbar \omega _{l,\bar{n}}\left( 1+\bar{n}+|\iota |\right) -\eta ^{2}/\left(
2m\omega _{l,\bar{n}}^{2}\right) $ in terms of the allowed frequencies $%
\omega _{l,\bar{n}}$. Since the eigenvalue equation (\ref{eq:eig_eq_VB_LH})
supports bound states for all $-\infty <\theta <\infty $ we realize that
something is amiss here too.

The potential-energy function for the fifth example is $V(r)=\eta
r+k/r+m\omega ^{2}r^{2}/2$ and the resulting eigenvalue equation reads
\begin{equation}
u^{\prime \prime }(s)+\frac{1}{s}u^{\prime }(s)-\frac{\iota ^{2}}{s^{2}}u(s)-%
\frac{\gamma }{s}u(s)-\theta su(s)-s^{2}u(s)+\delta u(s)=0.
\label{eq:eig_eq_VB_CLH}
\end{equation}
Also in this case the authors found allowed values of the oscillator
frequency that, as argued above, is unexpected.

Vit\'{o}ria and Belich\cite{VB19} argued that the solutions to equations (%
\ref{eq:eig_eq_VB_CH}), (\ref{eq:eig_eq_VB_LH}) and (\ref{eq:eig_eq_VB_CLH})
can be expressed in terms of the solutions to a biconfluent Heun equation;
however, they never used the mathematical properties of the latter equation
explicitly and tried the Frobenius method outlined below.

All the examples discussed by Vit\'{o}ria and Bakke\cite{VB19} are
particular cases of the eigenvalue equation
\begin{equation}
u^{\prime \prime }(x)+\frac{1}{x}u(x)-\frac{\gamma ^{2}}{x^{2}}u(x)-\frac{a}{%
x}u(x)-bxu(x)-x^{2}u(x)+Wu(x)=0,  \label{eq:eig_eq}
\end{equation}
where $\gamma $, $a$ and $b$ are real model parameters (notice that $\gamma $
has nothing to do with the parameter in equations (\ref{eq:eig_eq_VB_CH})
and (\ref{eq:eig_eq_VB_CLH})). This eigenvalue equation has square
integrable solutions
\begin{equation}
\int_{0}^{\infty }\left| u(x)\right| ^{2}x\,dx<\infty ,
\label{eq:bound_states}
\end{equation}
for all $-\infty <a,b<\infty $ for an infinite number of allowed values of $%
W(a,b)$. Such eigenvalues satisfy the Hellmann-Feynman theorem\cite{F39}
\begin{equation}
\frac{\partial W}{\partial a}=\left\langle \frac{1}{x}\right\rangle >0,\;%
\frac{\partial W}{\partial b}=\left\langle x\right\rangle >0.  \label{eq:HFT}
\end{equation}

In what follows we apply the Frobenius method to the eigenvalue equation (%
\ref{eq:eig_eq}) by means of the ansatz
\begin{equation}
u(x)=x^{s}\exp \left( -\frac{b}{2}x-\frac{x^{2}}{2}\right)
P(x),\;P(x)=\sum_{j=0}^{\infty }c_{j}x^{j},\;s=\left| \gamma \right| .
\label{eq:ansatz}
\end{equation}
The expansion coefficients $c_{j}$ satisfy the three-term recurrence
relation
\begin{eqnarray}
c_{j+2} &=&A_{j}c_{j+1}+B_{j}c_{j},\;j=-1,0,1,2,\ldots ,\;c_{-1}=0,\;c_{0}=1,
\nonumber \\
A_{j} &=&\frac{2a+b\left( 2j+2s+3\right) }{2\left( j+2\right) \left[
j+2\left( s+1\right) \right] },\;B_{j}=\frac{4\left( 2j+2s-W+2\right) -b^{2}%
}{4\left( j+2\right) \left[ j+2\left( s+1\right) \right] }.  \label{eq:TTRR}
\end{eqnarray}
If the truncation condition $c_{n+1}=c_{n+2}=0$, $c_{n}\neq 0$, $%
n=0,1,\ldots $, has physically acceptable solutions for $a$, $b$ and $W$
then we obtain exact eigenfunctions because $c_{j}=0$ for all $j>n$. This
truncation condition is equivalent to $B_{n}=0$, $c_{n+1}=0$ or
\begin{equation}
W_{s}^{(n)}=2\left( n+s+1\right) -\frac{b^{2}}{4},\;c_{n+1}(a,b)=0,
\label{eq:trunc_cond}
\end{equation}
where the second condition determines a relationship between the parameters $%
a$ and $b$. On setting $W=W_{s}^{(n)}$ the coefficient $B_{j}$ takes a
simpler form:
\begin{equation}
B_{j}=\frac{2\left( j-n\right) }{\left( j+2\right) \left[ j+2\left(
s+1\right) \right] }.
\end{equation}
Notice that the truncation condition does not provide all the solutions but
only those for which the parameters $a$ and $b$ exhibit certain relations.
The reason is that this problem is not exactly solvable, as Vit\'{o}ria and
Belich appear to believe, but quasi-exactly solvable or conditionally
solvable (see \cite{CDW00,AF20,F20b,F20c} and, in particular, the remarkable
review \cite{T16} and references therein for more details).

As a first example we consider the eigenvalue equation (\ref{eq:eig_eq})
with $b=0$ that is defined by the potential $V(a,x)=a/x+x^{2}$. In this case
$c_{n+1}(a,0)=0$ is a polynomial function of $a$ of degree $n+1$ and it can
be proved that all the roots $a_{s}^{(n,i)}$, $i=1,2,\ldots ,n+1$, are real%
\cite{CDW00,AF20}. Besides, $c_{n+1}(a,0)=a^{j_{n}}Q_{n}\left( a^{2}\right) $%
, where $j_{n}=0$ for $n$ odd, $j_{n}=1$ for $n$ even and $Q_{n}\left(
a^{2}\right) $ is a polynomial function of $a^{2}$ of degree $\left(
n+1\right) /2$ for $n$ odd or $n/2$ for $n$ even. For convenience we arrange
the roots so that $a_{s}^{(n,i)}>a_{s}^{(n,i+1)}$ and stress the point that
all of them correspond to the same eigenvalue $W_{s}^{(n,i)}=W_{s}^{(n)}$.
It is important to realize that the eigenvalue $W_{s}^{(n)}$ is common to a
set of different quantum-mechanical problems given by $V_{s}^{(n,i)}(x)=V%
\left( a_{s}^{(n,i)},x\right) $. Part of the authors' mistakes stem from
overlooking this obvious fact. For example, the eigenvalues $\mathcal{E}_{l,%
\bar{n}}$ obtained by them correspond to different quantum-mechanical
problems and are, consequently, meaningless. The polynomial solutions
\begin{equation}
u_{s}^{(n,i)}(x)=x^{s}\exp \left( -\frac{x^{2}}{2}\right)
P_{s}^{(n,i)}(x),\;P_{s}^{(n,i)}(x)=\sum_{j=0}^{n}c_{j,s}^{(n,i)}x^{j},\;s=%
\left| \gamma \right| ,  \label{eq:poly_sol_CH}
\end{equation}
share the same eigenvalue $W_{s}^{(n)}$ and also correspond to different
quantum-mechanical problems $V_{s}^{(n,i)}(x)$. Another part of the authors'
mistakes comes from the belief that these polynomial solutions are the only
square integrable eigenfunctions supported by equation (\ref{eq:eig_eq}).

The actual eigenvalues $W_{\nu ,s}(a)$, $\nu =0,1,\ldots $, $W_{\nu
,s}<W_{\nu +1,s}$, of equation (\ref{eq:eig_eq}) with $b=0$ are curves in
the $a-W$ plane. It follows from the Hellmann-Feynman theorem (\ref{eq:HFT})
that $\left( a_{s}^{(n,i)},W_{s}^{(n)}\right) $ is a point on the curve $%
W_{i-i,s}(a)$. In order to verify this fact we need the actual eigenvalues $%
W_{\nu ,s}$ that we have to obtain by means of a suitable approximate method
because the eigenvalue equation (\ref{eq:eig_eq}) is not exactly solvable%
\cite{AF20,T16}. Here, we resort to the well known Rayleigh-Ritz variational
method that is known to yield upper bounds to all the eigenvalues\cite{P68}
and, for simplicity, choose the non-orthogonal basis set of Gaussian
functions $\left\{ \varphi _{j,s}(x)=x^{s+j}\exp \left( -\frac{x^{2}}{2}%
\right) ,\;j=0,1,\ldots \right\} $.

\begin{figure}[tbp]
\begin{center}
\includegraphics[width=9cm]{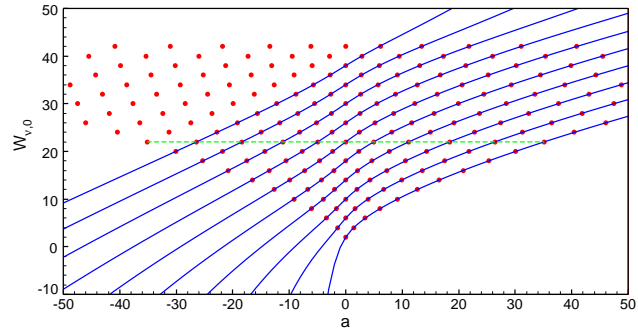}
\end{center}
\caption{Eigenvalues $W_0^{(n)}$ ($b=0$) from the truncation condition (red
points) and $W_{\nu,0}(a)$ obtained by means of the variational method (blue
lines)}
\label{Fig:Wb0g0}
\end{figure}

In order to make the variational calculations simpler we choose $s=0$ in
what follows. Figure~\ref{Fig:Wb0g0} shows several eigenvalues $W_{0}^{(n)}$
given by the truncation condition (red points) and the lowest actual
eigenvalues $W_{\nu ,0}(a)$ obtained from the variational method (blue
lines). We see that there are solutions to the eigenvalue equation (\ref
{eq:eig_eq}) for all values of $a$, that each $W_{\nu ,0}(a)$ is a
continuous function of $a$ that satisfies the Hellmann-Feynman theorem (\ref
{eq:HFT}) and that each pair $\left( a_{0}^{(n,i)},W_{0}^{(n)}\right) $ is a
point on those curves as argued above. This figure also shows the horizontal
line (green, dashed) for the mode $n=10$. Any vertical line starting from a
given value of $a$ will pass through no more that one red point. It means
that the truncation condition yields only one eigenvalue and just for a
particular model potential $V_{s}^{(n,i)}(x)$. An exception should be made
for the trivial case $a=0$ (harmonic oscillator) for which the truncation
method yields the whole spectrum. This particular case is the only one in
which the expansion coefficients satisfy a two-term recurrence relation and
the truncation approach is known to produce the actual spectrum of the
exactly-solvable quantum-mechanical model\cite{P68}. We realize that the
radial modes defined by Vit\'{o}ria and Belich have no physical meaning
unless one connects the points $\left( a_{s}^{(n,i)},W_{s}^{(n)}\right) $
properly. Since these authors were unaware of such connection they drew
nonsensical physical conclusions like the existence of allowed oscillator
frequencies.

As a second illustrative example we consider the eigenvalue equation (\ref
{eq:eig_eq}) with $a=0$ and denote $V(b,x)=bx+x^{2}$ the model potential. In
this case $c_{n+1}(0,b)$ is a polynomial function of $b$ of degree $n+1$
with roughly the same features discussed above for the preceding example.
There are also $n+1$ real roots $b_{s}^{(n,i)}$, $i=1,2,\ldots ,n+1$ that we
arrange in the same way: $b_{s}^{(n,i)}>b_{s}^{(n,i+1)}$. Each of them gives
rise to a model with the potential $V$ $_{s}^{(n,i)}(x)=V\left(
b_{s}^{(n,i)},x\right) $. The main difference with respect to the preceding
case is that the eigenvalues $W_{s}^{(n,i)}=2(n+s+1)-\frac{\left[
b_{s}^{(n,i)}\right] ^{2}}{4}$ lie on an inverted parabola instead of on an
horizontal straight line. As in the preceding example $\left(
b_{s}^{(n,i)},W_{s}^{(n,i)}\right) $ is a point on the curve $W_{i-1,s}(b)$.

\begin{figure}[tbp]
\begin{center}
\includegraphics[width=9cm]{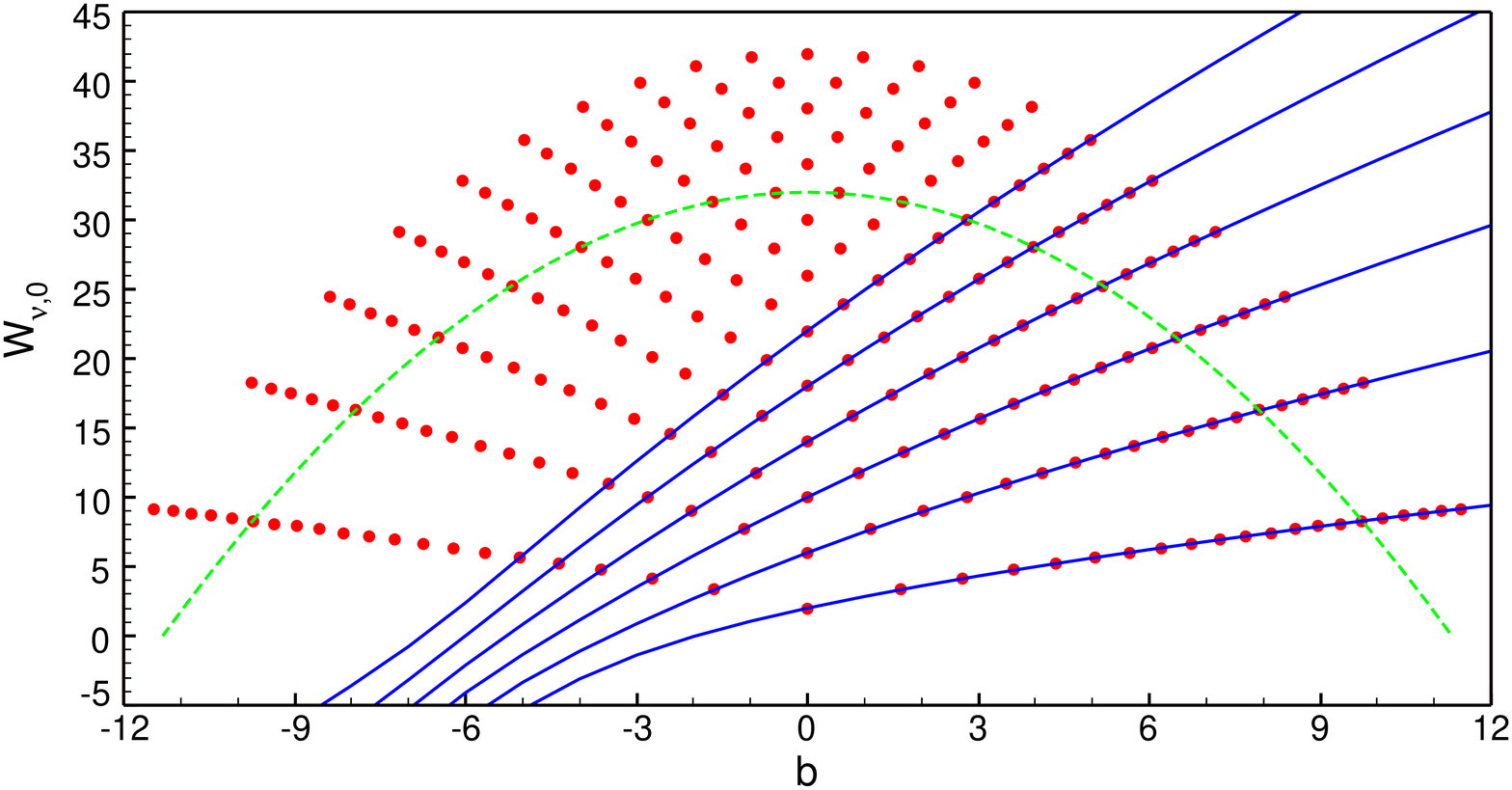}
\end{center}
\caption{Eigenvalues $W_0^{(n,i)}$ ($a=0$) from the truncation condition
(red points) and $W_{\nu,0}(b)$ obtained by means of the variational method
(blue lines)}
\label{Fig:Wa0g0}
\end{figure}

Figure~\ref{Fig:Wa0g0} shows some of the eigenvalues $W_{s}^{(n,i)}$ given
by the truncation condition (red points) and the lowest actual eigenvalues $%
W_{\nu ,s}(b)$ obtained by means of the variational method with the same
Gaussian basis set indicated above (blue lines). As in the preceding case,
we appreciate that the true eigenvalues of equation (\ref{eq:eig_eq}) with $%
a=0$ are continuous functions of $b$ that satisfy the Hellmann-Feynman
theorem (\ref{eq:HFT}). The allowed oscillator frequencies conjectured by Vit%
\'{o}ria and Belich\cite{VB19} are a consequence of misunderstanding the
meaning of the eigenvalues $W_{s}^{(n,i)}$ given by the truncation
condition. Since they failed to connect them properly they could not
understand that such eigenvalues are just points on the curves $W_{\nu ,s}(b)
$ as clearly shown in Figure~\ref{Fig:Wa0g0} for $s=0$. Notice that the
truncation condition yields at most one eigenvalue and just for a particular
model potential $V_{s}^{(n,i)}(x)$, exception being made for the trivial
case $b=0$ as argued above. Figure~\ref{Fig:Wa0g0} also shows the inverted
parabola (green, dashed line) for the mode $n=15$ that connects the points $%
\left( b_{0}^{(15,i)},W_{0}^{(15,i)}\right) $.

\begin{figure}[tbp]
\begin{center}
\includegraphics[width=9cm]{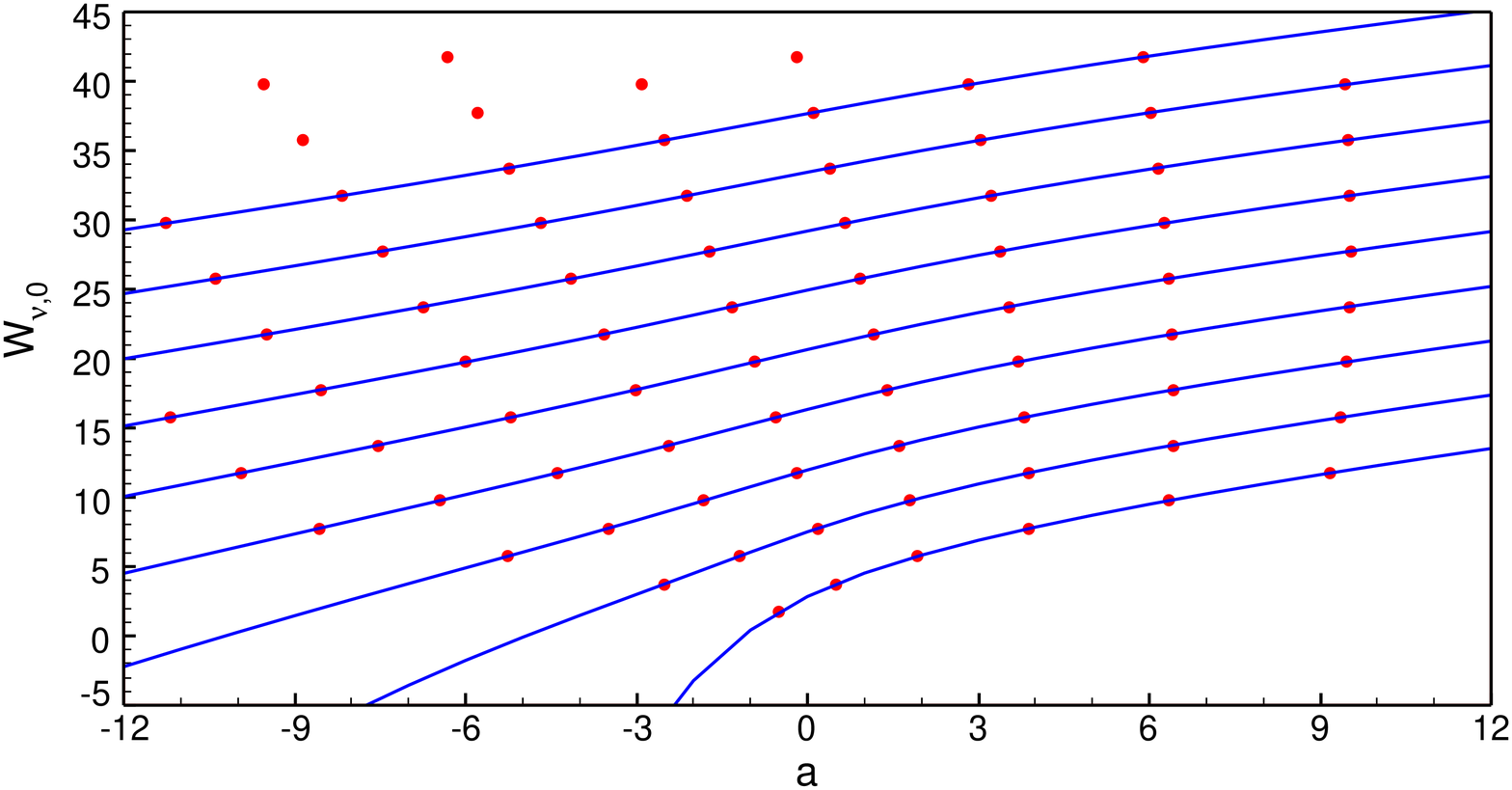}
\end{center}
\caption{Eigenvalues $W_0^{(n)}(a,1)$ from the truncation condition (red
points) and $W_{\nu,0}(a)$ obtained by means of the variational method (blue
lines)}
\label{Fig:Wb1g0}
\end{figure}

When $a\neq 0$ and $b\neq 0$ the true eigenvalues $W_{\nu ,s}(a,b)$ of
equation (\ref{eq:eig_eq}) are surfaces in the three dimensional space $%
\left( a,b,W\right) $. For simplicity we choose $b=1$ so that $W_{\nu
,s}(a,1)$ is a curve as before. Figure~\ref{Fig:Wb1g0} shows eigenvalues $%
W_{0}^{(n)}$ for $b=1$ and $s=0$ (red points) and variational results $%
W_{\nu ,0}(a,1)$ (blue lines). If we compare the distribution of the
eigenvalues $W_{0}^{(n)}$ for $b=0$ and $b=1$, given by the truncation
method, we appreciate that in the latter case the symmetry is lost and that
the exact results for the harmonic oscillator do not appear. In this case
the truncation method only yields one eigenvalue and just for some model
potentials given by particular values of $a$ (the roots of $c_{n+1}(a,1)=0$).

\textit{Summarizing}: It follows from present analysis that the eigenvalues $%
W_{\nu ,s}(a,b)$ of equation (\ref{eq:eig_eq}) are continuous functions of $%
-\infty <a<\infty $ and $-\infty <b<\infty $; therefore, the \textit{allowed}
or \textit{permitted} oscillator frequencies $\omega _{l,\bar{n}}$ were
fabricated by Vit\'{o}ria and Belich\cite{VB19} by means of a wrong
interpretation of the meaning of the roots provided by the truncation
method. The analytical eigenvalues presented by these authors are
meaningless because they correspond to different model potentials. More
precisely, the roots of the truncation condition are meaningless unless one
arranges and connects them in a suitable way as shown in present figures \ref
{Fig:Wb0g0} and \ref{Fig:Wa0g0}.

\end{document}